\documentclass[preprint,10pt]{aastex}
\newcommand{\grsim}{\gtrsim} 

\begin{document}
\title{Implications of Neutron Decoupling in Short Gamma Ray Bursts}

\author{Jason Pruet and Neal Dalal} \affil{Physics
Department, University of California, San Diego, CA 92093}

\begin{abstract}

Roughly half of the observed gamma-ray bursts (GRBs) may arise from
the shocking of an ultra-relativistic shell of protons with the
interstellar medium (ISM).  Any neutrons originally present in the GRB
fireball may, depending on the characteristics of the central engine,
dynamically decouple as the fireball accelerates.  This leads to
outflow consisting of separate fast proton and slow neutron
components. We derive detailed implications of neutron decoupling for
the observed lightcurves of short bursts.  We show that the collision
of a neutron decayed shell with a decelerating outer shell is expected
to result in an observable second peak in the GRB lightcurve.  There
may be substantial optical emission associated with such an
event, so the upcoming Swift satellite may be able to place
constraints on models for short bursts.  We also discuss interesting
inferences about central engine characteristics allowed by existing
BATSE data and a consideration of neutron decoupling.
\end{abstract}


\section{Introduction} 

GRB emission is widely thought to arise from the synchrotron cooling
of electrons shock-heated in collisions involving ultra-relativistic
shell(s) of particles. The collisions generating the shocks may be
between two adjacent shells of ejecta (the ``internal shock'' scenario
\citep{narpacpir,reesmesz}, or between an ultra-relativistic shell and
the ISM (the ``external shock'' scenario \citep{meszrees}). It has
been argued that long complex bursts are most naturally explained by
internal shocks \cite{piraninternal2} (see, however,
\cite{dermer}). Short, simple bursts, on the other hand, are well
explained by external shocks onto a homogeneous ISM.  In both
scenarios, some compact, as of yet unidentified central engine must
give rise to the ultra-relativistic flow.  Leading models for the
central engine include collapsars \citep{woosley}, and the coalescence
of a neutron star with another compact object \citep{schramm}.

It is a principal aim of GRB researchers to determine the nature of
the central engine. In this work we show that, within the context of
the external shock model for short bursts due to radiation-driven
fireballs, existing BATSE and upcoming HETE-II, Swift, and Glast data
on GRB lightcurves can be used make new and interesting inferences
about GRB central engine parameters.  In particular, we argue that the
observed properties of short bursts may be used to differentiate
between central engines emitting neutron rich ejecta, such as neutron
star mergers, and central engines emitting neutron poor ejecta.  This
is interesting because neutron star mergers are at present a favored
candidate for producing the timescales seen in the short class of
GRBs.  If, indeed short bursts arise from a different class of central
engines than long bursts, and do not give rise to afterglows
\cite{rotse,hurley} then the results presented here may allow unique
insights into the nature of these events.

The new inferences we discuss are obtained by considering the
dynamics of the neutron component of the fireball.  The basic idea is
that if strong neutron-proton scatterings become ineffective at
coupling neutrons to an accelerating radiation-driven gas, then these
neutrons will be left behind with a smaller average Lorentz factor
than the strongly Coulomb-coupled protons in the gas
\citep{derishev,fuller,bahcall}.  When this ``neutron decoupling''
occurs, the final fireball flow consists of two separate components: 
a fast proton shell and a slower
neutron shell.  When the neutrons in the slower shell decay they
can shock with the decelerating proton component and give rise to a
second, possibly distinct peak in the observed photon emission. This
process is schematically illustrated in figure 1. 

Roughly what is expected in this scenario is a first burst in the
gamma-ray regime followed by a second burst which can peak in the
x-ray, UV, or optical band. The first burst is a signature of the 
outer proton shell and the second a signature of the decayed neutron
shell. Typically the delay between the bursts is a few milliseconds 
to several seconds, depending in detail on the neutron richness of the 
outflow and other central engine properties. The upcoming Swift 
\footnote{http://swift.sonoma.edu} satellite,
with a fast response time and x-ray and optical capabilities, may be ideally
suited to look for evidence of neutron decoupling.

Observable effects of neutron decoupling for external
shocks were first considered by \cite{derishev0}. Here we analyze in
detail implications of this idea.  In that work 
Derishev et al.\ also proposed the possibility of
emission resulting from the decay of neutrons after arriving at the
outer proton 
shock radius. As we show below, it is not clear if
this case can be described in terms of the standard shocks discussed
in connection with GRBs, and an analysis of the emission in this case
has not been done. We therefore concentrate in this paper principally
on the case where the neutrons decay before shocking with the outer
shell, which is describable in terms of the standard theory of shocks.

In the next section we relate the observed timescales of short bursts
to the necessary conditions for a decoupled neutron component
to carry a substantial fraction of the fireball energy and decay before
colliding with the outer shell.  We also
derive an interesting relation between neutron decoupling and the
condition that the reverse shock in the outer shell is
relativistic. In \S3 we discuss the shock emission from
the collision between the decoupled shell of decayed neutrons with the
outer decelerating proton shell.  This is similar to the work of
\cite{pk} who consider the implications of the late time emission of a
slow baryon shell for the GRB afterglow. In \S4 we use
our results to draw some interesting conclusions about the properties
of GRB central engines based on existing BATSE data. The last section
is a summary and discussion of our results.

\section{Short bursts due to external shocks and neutron decoupling}

We first lay the groundwork for our discussion by presenting some
results from the theory describing the shocking of an
ultra-relativistic proton shell on the ISM. This problem has been
considered by a number of authors and we refer the reader to
\cite{piranreview} for a review.

Consider a relativistic shell expanding into the ISM. As the shell
sweeps up the ISM it decelerates and two shocks form, a forward shock
propagating into the ISM and a reverse shock propagating into the
shell.  The forward shock is generically relativistic. However, the
character of the reverse shock and the details of deceleration of the
fireball depend on the dimensionless parameter
\begin{equation}
\label{xi}
\xi=(l/\Delta)^{1/2}\gamma ^{-4/3}
\end{equation}
 \citep{sariandpiran}.  Here
$l=(3E/4\pi n_{\rm ISM}m_p c^2)^{1/3}$ is the Sedov length, $\Delta$
is the width of the shell as measured in the observers rest frame
({\it i.e.} the frame in which the fireball is moving at $\gamma$),
$E$ is the total energy in the fireball, $n_{\rm ISM}$ is the number
density of the surrounding ambient ISM, and $m_p$ is the proton mass.
If we write the energy of the shell in units of $10^{51}{\rm erg}$ as
$E_{51}$, the density of ISM in units of ${\rm cm^{-3}}$ as $n_{ISM}$,
the Lorentz factor of the shell in units of $10^3$ as $\gamma_{3}$,
and the width of the shell in units of $10^7{\rm cm}$ as $\Delta_7$,
then $\xi\approx 30 \gamma_3^{-4/3} ((E_{51}/n_{\rm ISM})^{1/3}
{\Delta_7}^{-1})^{1/2}$. If $\xi \ll 1$ the reverse shock is
relativistic and the energy conversion occurs principally in the
reverse shock, while if $\xi \gg 1$ the reverse shock is Newtonian and
energy conversion takes place principally in the forward
shock. Interestingly, when $\xi\grsim 1$ initially, the shell spreads
laterally as it expands and drives $\xi$ to unity before
shocking. When $\xi$ is driven to unity the reverse shock becomes
moderately relativistic, so that substantial energy conversion in the
reverse shock also occurs in this case. We therefore focus on the case
where $\xi \le 1$ at the time of substantial energy emission (see also
\cite{sarinarayanpiran}). Therefore, in the formulas that follow $\xi$
is taken to be unity if initially $\xi>1$.
 
The bulk of the observed emission occurs when the expanding shell
slows substantially. This occurs when the energy
imparted to the ISM is a sizable fraction of the initial energy in the
shell. For $\xi \ll 1$ this corresponds to the point at which
the reverse shock crosses the shell. In both cases 
($\xi\ll 1$ and $\xi$ driven to unity throuh spreading) 
the deceleration radius is written as
\citep{meszrees,sariandpiran}
\begin{equation}
\label{rdec}
r_{\rm dec}=l/\gamma^{2/3} \xi^{-1/2} \approx 5\cdot 10^{15} {\rm cm}
(E_{51} /n_{\rm ISM})^{1/3} \gamma_3^{-2/3}\xi^{-1/2},
\end{equation} 
which also serves to specify the dynamic timescale
for the deceleration, $\sim r_{\rm dec}/c$. The observed duration of
the burst is then
\begin{equation}
\label{tburst}
T_b\approx r_{\rm dec}/c \gamma_{3}(r_{\rm dec})^2\xi^{-3/2}=(0.2 {\rm
sec}) \gamma_3^{-8/3} (E_{51}/n_{\rm ISM})^{1/3}\xi^{-2}.
\end{equation}

The above equation is uncertain to within a factor of a few owing to
the decrease in $\gamma$ (deceleration) of the shell as it propagates.
The factor of 
$\gamma^2$ in the denominator arises because of the relativistic
motion of the shell towards the observer \citep{rees}.  The factor of
$\xi^{-3/2}$ arises because when the reverse shock is relativistic the
Lorentz factor has decreased by a factor of $\xi^{-3/4}$ by the time
the reverse shock crosses the shell. Note that $T_b$ is equal to the
width of the relativistic shell at $r_{\rm dec}$. For $\xi<1$ lateral
spreading of the shell is unimportant and $T_b$ is just equal to
the duration of emission from the central engine. 
With the above equations in hand we can discuss the
implications of observed timescales of short bursts for neutron
decoupling.

For both internal and external shocks substantial baryon Lorentz
factors are required. A variety of mechanisms have been proposed to
explain how the material acquires such a large kinetic energy.  In a
leading scenario, the fireball model,
the baryons begin roughly at rest as part of a plasma with a large
ratio of total energy to baryonic rest mass ($\eta\equiv
E/M$). Thermal pressure then drives the acceleration of this
plasma. Protons are strongly coupled to this high entropy gas via
Thompson drag. Neutrons, however, are effectively only coupled via
strong scatterings with protons and will eventually decouple. The
condition that decoupling occurs before the end of the acceleration
phase of the fireballs evolution is
\begin{equation}
\label{decouple}
\eta_3 > .3 (E_{51}Y_e/R_6\tau_{\rm dur})^{1/4}.
\end{equation}
Here $\eta_3=\eta/10^3$, $Y_e=n_p/(n_n+n_p)$ is the net number of
protons per baryon in the fireball, $R_6$ is the central engine radius
in units of $10^6{\rm cm}$ (this parameter determines the acceleration of the
fireball), and $\tau_{\rm dur}$ is the duration of emission of the GRB
central engine in seconds (so that $\Delta/c=\tau_{\rm dur}$). When
Eq. \ref{decouple} is satisfied the protons will keep accelerating
while the neutrons are left behind. (If Eq. \ref{decouple} is not
satisfied the protons and neutrons still decouple, but have the same
Lorentz factors because the fireball is no longer accelerating.) When
Eq. \ref{decouple} is satisfied and the neutrons dynamically decouple 
during the acceleration stage of the fireball's evolution, the fraction 
of initial energy going to the neutrons is 
\begin{equation}
\label{fn}
f_n\approx {(1-Y_e) \over 5} \left({Y_e E_{51} \over \eta_3 ^4
 R_6\tau_{\rm dur}}\right)^{1/3},
\end{equation}
and the final Lorentz factor of the protons and neutrons is given by
\begin{equation}
\label{gammap}
\gamma_{p,3} = \gamma_{n,3} {(1-f_n)(1-Y_e) \over f_n Y_e} = {\eta_3
\over Y_e}(1-f_n) .
\end{equation} 
(See \cite{derishev}, \cite{fuller},or \cite{bahcall} for a detailed
derivation of the above equations). In this equation $\gamma_{p,3}$ and
$\gamma_{n,3}$ are, respectively, the final Lorentz factors of the
proton and neutron shells in units of $10^3$.  Of course, if the
neutrons initially present in the fireball are going to shock and lead
to an observable photon signature they must first decay into protons.
Even after they have decayed we will still refer to the slower inner
shell as the ``neutron'' shell. We note that it was argued in
\cite{fuller} that in fireballs with very low initial $Y_e$, the
electron fraction will be driven to $Y_e\sim 0.05$ during neutron
decoupling.

A useful relation, derivable from Eqs. \ref{fn} and \ref{gammap}, and
valid when neutron decoupling occurs, is
$\gamma_{p,3}^{4/3}=(1/5)(1-f_n)^{1/3}(\gamma_p/\gamma_n)
(E_{51}/r_6\tau_{\rm dur})^{1/3}$. Using this relation and Eq. \ref{xi}, 
note
that there is a very interesting connection between the conditions in
the neutron decoupled fireball and the initial value of the 
parameter $\xi$,
\begin{equation}
\xi=3{\gamma_n \over \gamma_p}
\label{neatxi}
\left({r_6^2 \over \tau_{\rm dur} n_{\rm ISM} E_{51}
(1-f_n)}\right)^{1/6}.
\end{equation}
This implies that for a given $\xi$, neutron decoupling occurs in the
progenitor fireball unless $n_{\rm ISM} E_{51} \tau_{\rm dur}
Y_e/r_6^2>(3)^6 \xi^{-6}$. Therefore, {\it a strongly relativistic
reverse shock ($\xi \ll 1$) implies that neutron decoupling has occurred in the
progenitor fireball for essentially all reasonable fireball
parameters.  }

We are interested in the conditions under which a slow neutron shell
decays and shocks with the outer proton shell. In order for this to
occur, and in order for observable emission to result, the following
conditions must be met: i) neutron decoupling occurs and leads to a
final ratio of proton to neutron Lorentz factors of
$\gamma_{p,3}/\gamma_{n,3}>\alpha$. Here $\alpha$ determines the
strength of neutron decoupling and is chosen so that the emission from
the two peaks is distinguishable. ii) the fraction of energy going to
the neutrons is non-negligible (for definiteness we impose $f_n>0.2$),
and iii) the neutrons decay before colliding and shocking with the outer
decelerating shell.  We denote with a subscript $p$
properties of the observed emission from the (faster) proton shell (so
that $T_{b,p}$ is the observed duration of the first peak), and with a
subscript $n$ properties of the observed emission from the decayed
neutron shell. When $\xi>1$, the expression for the duration of the
emission from the proton shell can be written in the useful form
\begin{equation} 
\label{usefultbp}
T_{b,p}=5(1-f_n)^{-1/3}\left(\gamma_n\over \gamma_p\right)^2
\left(\left(r_6 \tau_{\rm dur}\right)^2 \over E_{51} n_{\rm
ISM}\right)^{1/3}
\end{equation}

Eq. \ref{usefultbp} allows us to relate the observed proton shell
burst duration to the condition that the neutrons strongly decouple
and carry a substantial fraction of the fireball energy (conditions
(i) and (ii) above). Noting that
$f_n=(1-Y_e)/(1+(\gamma_p/\gamma_n-1)Y_e)$ we see that 
$\gamma_p/\gamma_n>\alpha$ and $f_n>0.2$ when

\begin{equation}
\label{tltgt}
{5\over 16} \left( Y_e \over 1-Y_e\right)^2 <T_{b,p}\left({ E_{51}
n_{\rm ISM} \over (r_6 \tau_{\rm dur})}\right)^{1/3} < 5 \alpha^{-2}
\left( {1+(\alpha-1)Y_e \over \alpha Y_e }\right)^{1/3}.
\end{equation}
For given central engine parameters this range is large if $Y_e$ is
small, and vice versa. This is because when $Y_e$ is small ({\it i.e.}
the fireball material is neutron rich) $\gamma_p /
\gamma_n$ can be large while still leaving a substantial portion of
the energy in the neutron component.  Eq. \ref{tltgt} only applies for
$\xi > 1$ initially and driven to unity through spreading. Note that
when $\xi < 1$, $T_{b,p}= \tau_{\rm dur}$, and while condition (i) is
easily satisfied, it is not possible to express condition
(ii) in terms of observables. 
This is because the burst duration has no relation to
$\gamma_p/\gamma_n$ when $\xi<1$.

In order to determine the properties of the burst arising from the
neutron shell and also to determine whether or not the neutrons decay
before colliding with the outer shell we need to specify how the outer
shell slows with time.  This requires knowing whether the evolution of
the outer shell is adiabatic or radiative. Adiabatic evolution refers
to the case where the energy generated in shocks with the ISM is not
radiated away ({\it i.e.}  most of the post shock energy is not in the
electrons), or is radiated away on a timescale slow compared to the
dynamic time of the fireball. The evolution is radiative if all the
energy generated in the shocks with the ISM is effectively
instantaneously lost from the system. This occurs when the fraction of
post-shock thermal energy is principally in the electrons
($\epsilon_e\sim 1$) and the electron energy is radiated away quickly.

A common assumption regarding the fraction of post-shock energy going
to the electrons is that $\epsilon_e$ is in the range $0.1-0.3$, {\it
i.e.}  significantly less than unity. In this case the outer proton
shell initially slows approximately adiabatically regardless of the
cooling time for the electrons.  Because a radiative evolution is not
ruled out, and to bracket the range of possible behaviors, we will
also note results for the radiative case where appropriate. The true 
behavior will be somewhere inbetween. We will
see that the characteristic timescales of the burst from the neutron
shell are not very sensitive to the choice of adiabatic or radiative
evolution of the outer shell. The spectral characteristics of the burst
from the neutron shell, on the other hand, are more sensitive to the
conditions in the outer proton shell when the two shells collide.

For the case of a Newtonian reverse shock ($\xi\sim 1$),
\citet{katzpiran} give the following analytic approximation to
describe the slowing of the shell:

\begin{equation}
\label{gammapoft}
\gamma_p(t)={\gamma_p(t=0) \over 2} \left({r_{\rm dec} \over c
t}\right)^{3/2}.\end{equation}
For radiative evolution the exponent $3/2$ above becomes 
$3$. The time $t$ appearing in Eq. \ref{gammapoft} above and also in 
Eqs. \ref{goftxilt1} below is the time as measured by an observer at rest 
with respect to the central engine.

For the case where the reverse shock is relativistic the evolution is
more complicated. We follow \cite{sari} in describing the evolution of
the proton shell by a broken power law.

\begin{eqnarray}
\label{goftxilt1}
\gamma_p(t)&=&{\gamma_p(t=0) \xi^{3/4} \over 2} \left({r_{\rm dec}
\over c t}\right)^{1/2}\qquad\hbox{for   }\xi^{3/2}r_{\rm dec}<r<r_{\rm dec} \\
\gamma_p(t)&=&{\gamma_p(t=0) \xi^{3/4} \over 2} \left({r_{\rm dec}
\over c t}\right)^{3/2}\qquad\hbox{for   }r>r_{\rm dec}
\end{eqnarray}

Again the exponent $3/2$ describing the later 
evolution would be $3$ for a radiative evolution. 
When the reverse shock is relativistic a
collision between the neutron and proton shells could occur before
$r=r_{\rm dec}$. This requires $\gamma_p(t=0)\xi^{3/4}\ll\gamma_n$,
which is difficult to satisfy unless $\tau_{\rm dur}$ is very large
and which also implies that the emission from the proton and neutron
shells would not be well separated. In what follows we therefore
concentrate on the case where $\gamma_p(t=0)\xi^{3/4}\ge
\gamma_n$. Note that when $\gamma_p(t=0)\xi^{3/4}\ge \gamma_n$,
spreading is important for the neutron shell because $t_{\rm
collide}/\gamma_n^2>\tau_{\rm dur}$.

With the above prescription for the evolution of the proton shell the
two shells collide when

\begin{equation}
\label{gammarel}
(\gamma_p(t)/\gamma_n)^2=1/4
\end{equation}
independent of $\xi$
and at a time
\begin{equation}
\label{tcollide}
t_{\rm collide} \approx \left({\gamma_p
 \xi^{3/4}\over \gamma_n}\right)^{2/3} r_{\rm dec}/c.
\end{equation}

The fraction of neutrons decaying by this time is $f_{\rm decay}
\approx 2.5 (E_{51}(1-f_n))^{-1/12}n_{\rm ISM}^{-1/3}(r_6\tau_{\rm dur})^{5/12}
(\gamma_p/\gamma_n)^{5/12}$.  The condition that $f_{\rm decay}$
is substantial ($f_{\rm decay}>0.5$) is 
\begin{equation}
\label{neutrondecay}
r_6 \tau_{\rm dur} (\gamma_p/\gamma_n)> 0.02 n_{\rm ISM}^{4/5}
(E_{51}(1-f_n))^{1/5}.
\end{equation} 

This equation is interesting because the fraction of neutrons decaying
depends so weakly on all of the fireball parameters except $r_6
\tau_{\rm dur}$ and $\gamma_p/\gamma_n$. For comparison, we note that 
when  the evolution of the 
outer shell is radiative rather than adiabatic the condition that 
$f_{\rm decay}$ is substantial becomes
\begin{equation}
\label{neutrondecay2}
r_6 \tau_{\rm dur} > 0.04 \xi^{3/5}n_{\rm ISM}^{4/5}
(E_{51}(1-f_n)\gamma_n/\gamma_p)^{1/5}.
\end{equation} 

In the next section we turn to a description of the emission resulting
from the shocking of the slower inner ``neutron'' shell once it
collides with the decelerating outer proton shell. First, though, we
note that \citet{derishev0} have
suggested that observable emission may
result from the case where the neutrons decay after colliding with the
proton shell.  A quantitative assessment of
this suggestion is not possible within the context of the standard
picture of one body impinging and shocking on another. To see this,
note that as the neutrons travel and decay they will decay on top of
some ISM rest mass. In a time $dt$ as measured by an observer at rest
in the central engine rest frame a fraction of energy $dE_{\rm n
\rightarrow p} \sim E_n dt/(\tau_n \gamma_n)$ appears in the form of
protons from neutron decay. In this same time $dt$, the neutron shell
will sweep over an ISM rest mass of $d{\rm M_{ISM}}=r^2c\,dt\,r_{\rm
dec}^{-3} E_p\gamma_p^{-2}\xi^{-3/2}$ (this is obtained by noting that
the ISM rest mass within $r_{\rm dec}^3$ is
$\gamma_p^{-2}\xi^{-3/2}E_p$).  A description of the ensuing process
as a sweeping up a small amount of material and then a shocking is
only possible if $\gamma_n^2d{\rm M_{ISM}} \ll dE_{\rm n\rightarrow
p}$, or equivalently if
\begin{equation}
\label{w}
\gamma_n\left({c\tau_n r^2 \over r_{\rm dec}^3}\right) \left({\gamma_n
\over \gamma_p}\right)^2\xi^{-3/2}\ll 1
\end{equation}
Because $\tau_n c >r_{\rm dec}$ is the condition that neutrons haven't
decayed by $r_{\rm dec}$, Eq. \ref{w} is in general not satisfied.  
This does not preclude substantial, observable emission from other
processes, for example plasma instabilities.
Further study may provide interesting insights.

\section{Emission from the ``neutron'' shell}

When the conditions presented above are satisfied, neutrons in the
inner shell decouple from the proton shell, carry an
energy comparable to the energy in the proton shell, and 
decay by the time they collide with the outer proton shell. This
collision occurs at approximately the time $t_{\rm collide}$ given
above and will generate a forward and a reverse shock. The characteristics
of these shocks and the resultant emission depend on the
structure of the outer shell at the time of the collision. We will consider 
two limiting approximations to this structure. 

In the first approximation that we consider, the outer shell is assumed to 
have relaxed to a Blandford-McKee self similar solution. Also, the outer shell 
is assumed to be hot (at least in the proton component), so that the 
enthalpy in the outer shell is much larger than the rest mass energy density.
\cite{kobpirsari} studied numerically the evolution of external 
shocks and showed that the Blandford-Mckee self similar solution is a 
good approximation once the shock has reached a radius greater than
$\sim 1.4-1.9 r_{\rm dec}$, with the exact number depending on whether
or not the reverse shock is relativistic. Therefore, for reasonable fireball
parameters the collision between the inner and outer shells may occur
while the outer shell is still relaxing to the self similar solution.
A numerical solution of the evolution of the inner and outer shells
would be needed to obtain the expected lightcurves in this case.

The expected emission in this first aproximation has been worked out in 
detail by \cite{pk} and we draw on their results. When the inner cold 
shell collides with the outer hot shell a weak forward shock results. 
The effect on the emission from the outer shell is a modest increase in 
the total luminosity and little change in the spectrum. The reverse shock
propagating into the inner shell is strong and mildly relativistic. The 
characteristic frequency for the emission from the inner shell is 

\begin{equation}
\label{charfreq}
(h\nu_{syn})_{|\gamma_{e,min}}\approx 1 {\rm keV} \gamma_{n,3}^{5/2}
\epsilon_e^2 \epsilon_b^{1/2} n_{\rm ISM}^{1/2} (E_p/E_n)^{3/2}
\end{equation}

This expression should be taken as somewhat approximate because a
numerical study of the evolution of the reverse shock propagating into
the neutron shell is needed for an accurate determination of the
characteristic frequency.  The flux at the characteristic frequency is
larger by a factor $\sim(\gamma_n E_p/E_n)^{5/3}$ than the flux from
the outer shell at the same frequency \citep{pk}.  If the observed
emission from the proton shell arose from the reverse shock in the
proton shell (which might occur if the emission from the forward shock
occurs at too high a frequency to be observed by BATSE), then the
characteristic frequency in the first and second peaks is similar. If, on
the other hand, the dominant emission in the first peak arose from the
forward shock propagating into the ISM, the first peak will have a
much higher average energy than the second peak.  The future GLAST
mission\footnote{http://glast.gsfc.nasa.gov} may therefore provide
useful insights into this problem.

The spectrum of emission from the inner shell depends on whether or
not the typical post-shock electron cools within a dynamical
timescale.  The thermal Lorentz factor of an electron which just cools
on a hydrodynamic timescale is given by $\gamma_{e,c}=3m_e
c/(4\sigma_{\rm T} U_B \gamma_n t_{\rm hyd}) \approx 3(E_p/E_n)(1/n_{\rm
ISM}\gamma_{n,3}^3\epsilon_B t)$. Here $U_B$ is the magnetic field
energy density in the inner neutron shell, $\sigma_{\rm T}$ is the
Thomson cross section, $t_{\rm hyd}$ is the observed hydrodynamic
timescale for the shell (approximately $r/c \gamma_n^2$), and $t$ is
the observed time in seconds.  Because the thermal Lorentz factor
($\gamma_{e,{\rm therm}}$) for the average electron in the
post-shocked inner shell is typically of order a few hundred or
higher, we see that $\gamma_{e,c}\lesssim \gamma_{e,{\rm therm}}$
during the first few seconds.  This means that we may approximate the
electrons as fast cooling.  In this case the flux from the neutron
shell is proportional to $\nu^{-p/2}$ for frequencies above the
characteristic frequency . Here $p$ is the power law index
characterizing the electron distribution in the shock (typically
$p\sim 2.4$).  Even though the characteristic frequency for the inner
shell is low, significant emission can occur in the tens to hundreds
of keV range if the index of the power law characterizing the
post-shock electron distribution is close to 2. 

The second approximation we
consider is the case where essentially all of the shock energy
generated in the outer shell goes into the electrons and is radiated
away rapidly. In this case the two shells collide when they are cold
and the collision will generate comparable forward and reverse
shocks. The emission will be similar to that described in the first
approximation discussed above. However, the characteristic frequency
from these shocks can be a factor of tens to hundreds larger in this case
because of the larger relative Lorentz factor of the shells at the
time of the collision and because of the smaller enthalpy of the outer
shell at the time of collision.

Interestingly, the signature from a neutron decayed shell
can be quite similar to what is expected for the case where, following
internal shocks, a relativistic shell collides with the ISM
\citep{norotse}.  For example, for GRB 970228 one might interpret the
first peak as arising from the forward shock occurring when a fast
proton shell collides with the ISM, and the second peak as arising
from the collision of a neutron decayed shell with the outer shell.

When the neutron shell leads to an observable peak, this second peak
will have a characteristic duration
\begin{equation}
\label{tbn}
T_{b,n}\approx t_{\rm collide}/{\gamma_n}^2\approx T_{b,p}
\left(\xi^{3/4} \gamma_p /\gamma_n \right)^{8/3}
\end{equation}
and the second peak will be separated from the first peak by an
observed time
\begin{equation}
\label{deltat}
\delta t \approx \left(T_{b,p}  \right) (\gamma_p \xi^{3/4}/ \gamma_n)^2
(( \gamma_p \xi^{3/4}/ \gamma_n)^{2/3}-1).
\end{equation}

Although the results for the cases where $\xi<1$ and where $\xi>1$
initially can be written in similar forms, there is an important
difference between these two cases.  In the newtonian reverse shock
case ($\xi$ is driven to unity through spreading), $T_{b,n}$ and
$\delta t$ have a relatively strong dependence on
$\gamma_p/\gamma_n$. Therefore, for the newtonian case, one could have
$\gamma_p/\gamma_n=10$, for example, leading to a second peak
separated from the first by more than 100 first peak
durations. However, for $\xi<1$, $\gamma_p \xi^{3/4} / \gamma_n
\approx 2 (\gamma_p / \gamma_n)^{1/4} (r_6^2/{\tau_{\rm dur} n_{\rm
ISM} E_{51} (1-f_n)})^{1/8}$ and the dependence of $T_{b,n}$ and
$\delta t$ on $\gamma_p/\gamma_n$ is weak.

\section{Information about central engine parameters from observed 
lightcurves of short bursts}

In this section we illustrate how the observed temporal
characteristics of short, simple GRBs may be used to determine
properties of the burster central engine.
We emphasize again that we are supposing
short bursts to arise from external shocks.  We also assume that
thermal pressure (a fireball), and not for example magnetic fields,
drives the acceleration of the baryons. Both are debated assumptions.

There are two types of short bursts amenable to our analysis.  The
first class is the set of bursts with single peaks of emission; we
show below how the absence of a second peak leads to interesting
constraints.  The second class is composed of bursts with two or more
peaks.  These bursts can be used to infer central engine properties by
attributing a portion of the burst to a neutron decayed shell.  This
second method is attractive because in principle it places strong
constraints on the central engine parameters and the electron fraction
in the outflow. However, because many short bursts have structures too
complicated to be explained within the context of the neutron
decoupling scenario (how does one get three peaks for example), this
argument could only be convincing in a statistical sense.

We first examine the constraints on central engine parameters implied
by single peaked events. A basic confounding issue with this analysis
is the uncertainty in the parameter $\xi$, which describes the
strength of reverse shock in the outer proton shell.  If $\xi>1$, then
the absence of a second peak implies that one or both of Eqs.
\ref{tltgt},\ref{neutrondecay} were violated. The resultant
constraints are illustrated in figure 1. If $\xi<1$, it is in general
easy to satisfy the neutron decoupling condition. However, as the
burst duration is not related to $\gamma_p/\gamma_n$ in this case, the
absence of a second peak only gives weak constraints on the final
Lorentz factor $\eta$ of the fireball.  

As a definite example, consider a neutron star merger model characterized by
electron fraction $Y_e\sim 0.1$, $r_6\approx 1$, and a duration of a
few orbital periods ($\tau_{\rm dur}\sim 0.01 s$). Now, 
all bursts with durations greater than $\tau_{\rm dur}$ (a few tens of
msec) arise from fireballs with $\xi>1$. If the neutron decay condition
(Eq. \ref{neutrondecay}) is satisfied, then one expects the presence
of a second peak for all bursts with duration $\tau_{\rm
dur}<T_{b,p}(E_{51}n_{\rm ISM})^{1/3}\le 0.5 s.$ (Here we have taken
$\alpha=2$ as the criterion that the first and second peaks are
distinguishable.) For this range of burst durations it is readily seen
that the neutron decay condition (Eq. \ref{neutrondecay}) is satisfied
for $n_{\rm ISM}\sim1$ when the evolution of the outer shell is
approximately adiabatic. Therefore, a second peak in the GRB
lightcurve for these events is expected.

We now consider short bursts with multiple peaks.  A difficulty here
is that the majority of bursts display complex time structure with
many peaks.  For these bursts an
inhomogeneous ISM or some other mechanism must be invoked if they are
to be explained by an external shock model. An analysis of
implications of neutron decoupling for, {\it e.g.} an inhomogeneous
ISM, would have to be done in order to look for correlations
characteristic of neutron decoupling in these events.  Here for
simplicity we will focus on that subclass of bursts that display only
two peaks. Of course, second peaks in these events may arise from the
same mechanism that generates multiple peaks in more complicated
bursts, and not from neutron decoupling.

The general features of a two peaked burst due to neutron decoupling
are given in Eqs. \ref{tbn} and \ref{deltat}. A first proton shell
peak is followed by a longer neutron shell peak, with an inter-peak
duration somewhat shorter than the neutron peak duration. When $\xi>1$
initially and driven to unity through spreading, the ratio of first to
second peak widths is a direct measure of $\gamma_p/\gamma_n$. In
turn, $\gamma_p/\gamma_n$ gives the relative energies of the two
shells for a given $Y_e$. In particular, a large inferred
$\gamma_p/\gamma_n$ and comparable energies in the two peaks implies a
low $Y_e$.  A signature of low $Y_e$ environments might be the
presence of a population of bursts with $T_{b,n}/T_{b,p}\gtrsim 100$.

For $\xi<1$ it is difficult to make interesting inferences about
central engine parameters from two peaked bursts. This is because i)
as noted above the neutron decoupling condition is in general
satisfied for $\xi<1$, ii) the requirement that $f_n$ is
non-negligible only leads to weak constraints on the final Lorentz
factor $\eta$ of the
central engine, and iii) the neutron decay condition is principally
sensitive to $\tau_{\rm dur}$ which is directly measured for $\xi<1$. In
addition, the neutron decay condition depends on $\xi$, which is not
measured. Lastly, the weakness of the dependence of $T_{b,n}/T_{b,p}$
on the fireball parameters means that a ratio of first to second peak
widths of order unity is compatible with a broad range of conditions.

To make the connection between the work presented here and
observations of GRBs we display in figure 2 a plot of $\delta
t/T_{b,p}$ versus $T_{b,n}/T_{b,p}$ for a sample of bursts that
display only two 
peaks in BATSE energy channel 2. A proper analysis of these bursts
would take into consideration the fact that a burst which displays
only two peaks in a given energy channel may have more or fewer peaks
in different energy channels. The fits we use were done by Andrew Lee
\citep{lee1,lee2}.  Peaks are described by a function of the form
$I(t)=A\exp(-|(t-t_{\rm max})/\sigma_{r,d}|^{\nu})$. Here $I(t)$ is the
observed intensity, $t_{\rm max}$ is the time at which the peak
attains its maximum, $\sigma_r$ and $\sigma_d$ are the peak rise and
decay times respectively, and $\nu$ characterizes the shape of the
peak. We have followed \cite{lee2} in taking $\delta t$ to be the
difference between $t_{\rm max}$ for the two peaks, and in
approximating each peak width as
$T_{b}=(\sigma_r+\sigma_d)(-\ln(1/2))^{1/\nu}$.

A number of the bursts in figure 2 are compatible with the neutron
decoupling scenario presented here. If one attributes the cluster of
bursts with $\delta t/T_{b,p}\sim T_{b,n}/T_{b,p}\sim{\rm a\,few}$ as
arising from neutron decoupling, then either $\xi\sim 1$ and
$\gamma_p/\gamma_n\sim{\rm a\,few}$, or $\xi<1$ and
$\gamma_p/\gamma_n$ can be quite large if the outflowing material is
neutron rich.  Only the burst at $(\delta
t/T_{b,p},T_{b,n}/T_{b,p})=(130,192)$ might provide clear evidence for
pronounced neutron decoupling and low $Y_e$. However, because $\xi$ is
proportional to $\gamma_p/\gamma_n$ for a given set of central engine
parameters, the absence of a population of bursts with large
$T_{b,n}/T_{b,p}$ does not provide evidence against low $Y_e$ central
engines.  A natural explanation for the absence of such bursts for low
$Y_e$ central engine models is that when $\gamma_p/\gamma_n$ is large,
the reverse shock occurring when the outer proton shell collides with
the ISM is relativistic.

\subsection{Neutron decoupling and precursors}

Perhaps the most natural place to look for evidence of neutron
decoupling is in those bursts identified to have precursors. Roughly,
a precursor is a peak in the GRB lightcurve preceding the main
emission and separated from the main emission by a period where the
flux is dominated by the background.  \cite{koshut} studied in detail
24 bursts ($\sim 3\%$ of their total sample) that satisfied their
definition of having a precursor. Several of these 24 bursts are
consistent with the neutron decoupling picture presented here and
roughly half of them have $T_{b,n}/T_{b,p}\approx 10$. \cite{koshut}
also find a significant correlation between $T_{b,n}$ and $T_{b,p}$,
which is consistent with an explanation in terms of neutron
decoupling.  Most of the bursts they examine have several times more
emission in the second peak than in the first (although there is a
selection effect: they defined a precursor as an event having a
smaller peak count rate than the main emission). Within the context of
the neutron decoupling picture these precursor events might be
interpreted as evidence for a large neutron fraction.

\section{Conclusions}

Neutrons play an interesting role in relativistic
fireballs: the fraction of neutrons in the initial fireball may be a
direct indication of weak physics or other properties of the GRB
central engine \citep{pruetwinds}, the strong interaction cross section
is just right for neutrons to dynamically decouple during the
acceleration phase of a fireballs evolution, and the free neutron
lifetime allows for the possibility of neutrons decaying and shocking
with the outer proton shell when neutron decoupling occurs. Here we 
have explored in detail some implications of this last possibility. 

Somewhat fortuitously, an observable photon signature of the neutron
component is expected for a broad range of central engine
parameters. This signature arises from the reverse shock propagating
into the slower neutron decayed shell when it collides with the outer
proton shell and is characterized by a second peak in the GRB
lightcurve.  The characteristic frequency of this second peak is
typically in the few keV or lower range.  We have derived the relation
between the properties of these two peaks and the central engine
parameters.  The spectral and temporal correlations characteristic of
this second peak may be looked for in statistical studies of GRB
lightcurves. In principal, such studies could infer the electron
fraction in the GRB progenitor fireball. Low electron fractions, which
are thought to occur in neutron star-neutron star mergers, might
evidence themselves in a population of two peaked bursts with an
interpeak separation hundreds of times longer than the first peak
duration. The population of bursts which contain precursors and which 
are characterized by stronger emission in the second peak than in the
first may also be evidence for low electron fractions. Lastly, the upcoming
Swift satellite offers a promising opportunity to search for evidence of 
neutron decoupling and infer central engine properties.

\section{Acknowledgments}
\acknowledgments We thank Jim Wilson and Jay Salmonson for many useful
insights.  We are also indebted to Andrew Lee for providing us with
the GRB fit parameters used in his papers.  This work was partially
supported by NSF Grant PHY-098-00980 at UCSD and an IGPP mini-grant at
UCSD. ND was supported by the Dept.\ of Energy under grant
DOE-FG03-97-ER 40546 and a grant from the ARCS Foundation.

\begin{figure}
\centerline{\includegraphics[height=3in]{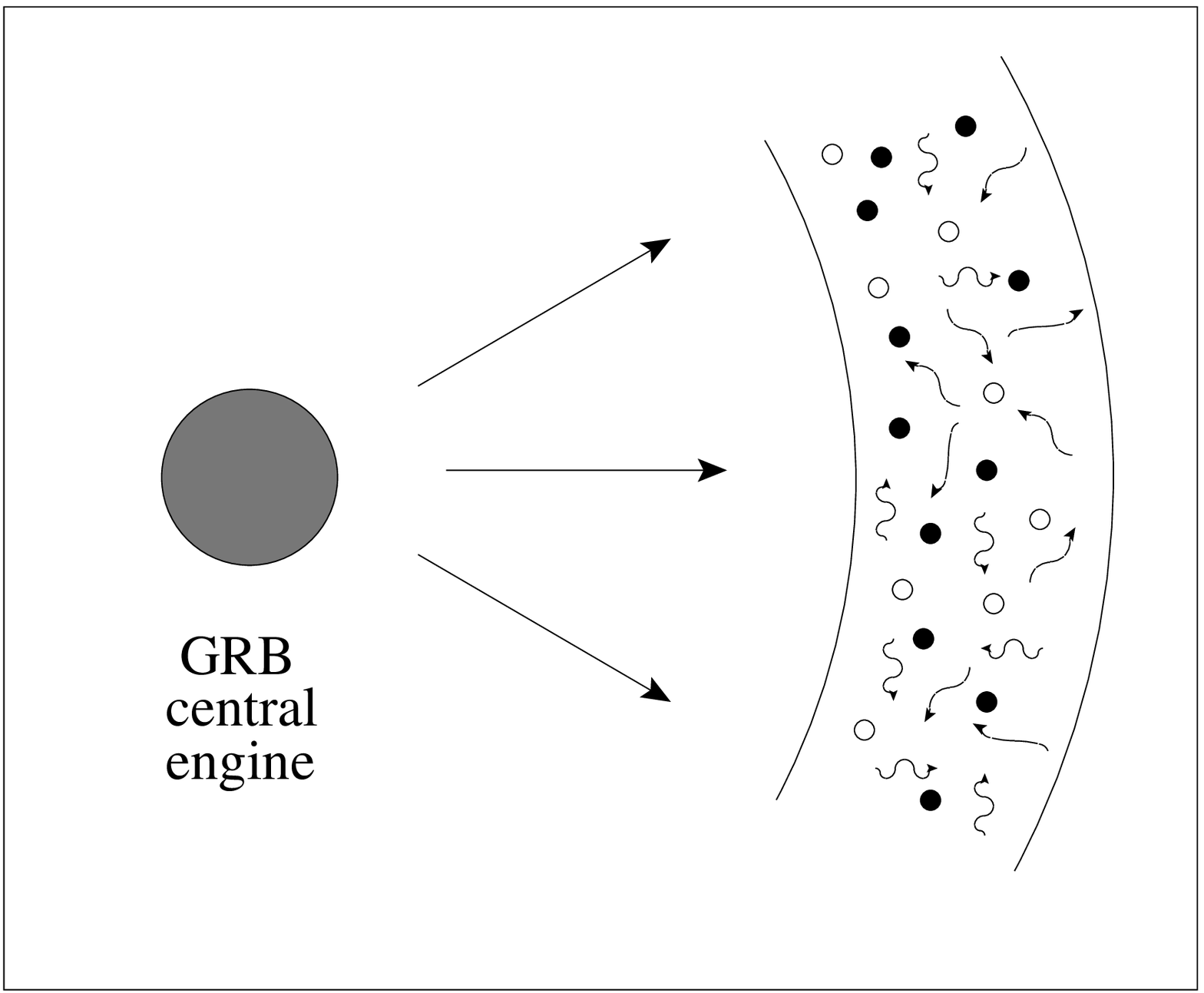}
\includegraphics[height=3in]{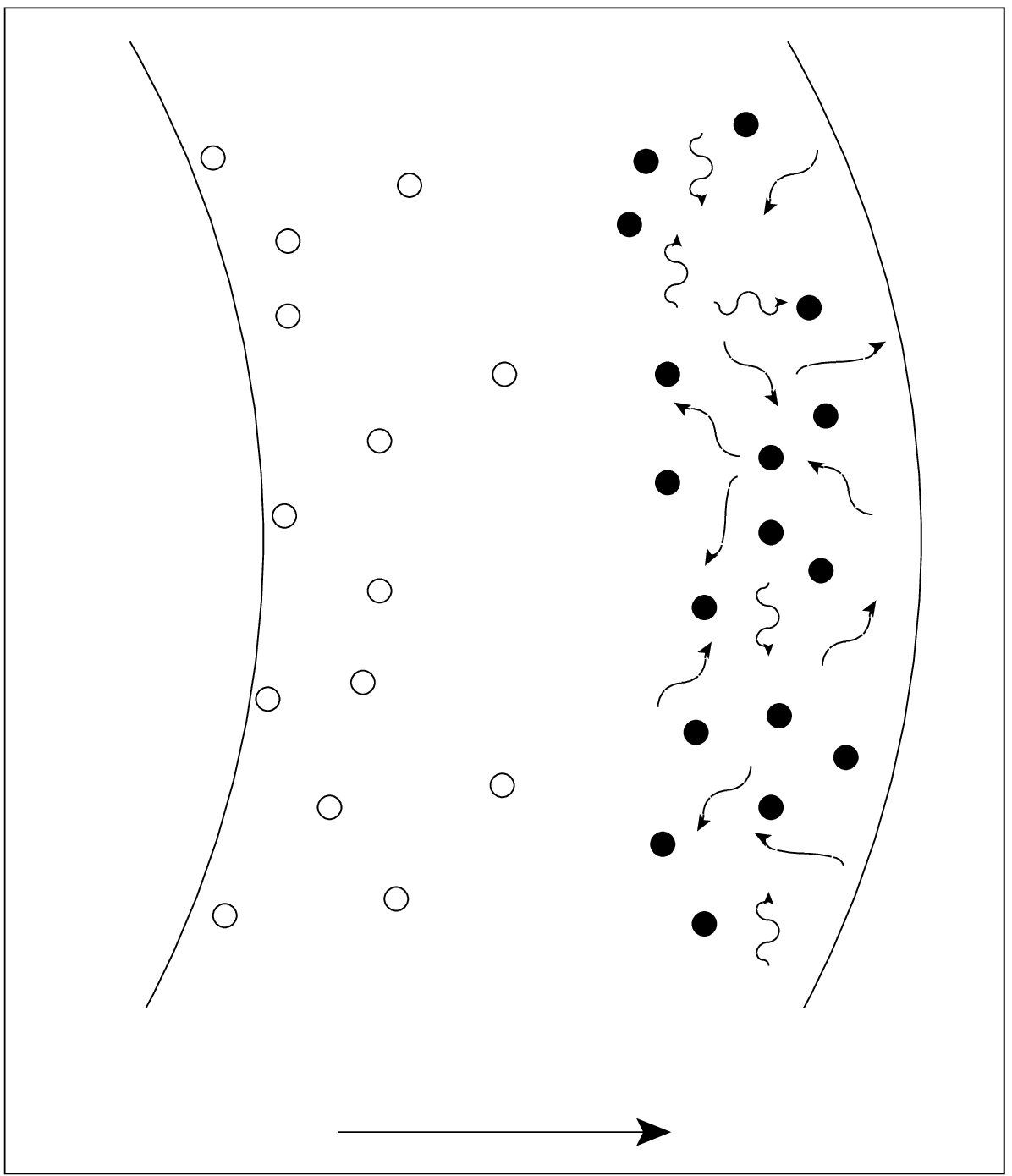}}
\centerline{\includegraphics[height=3in]{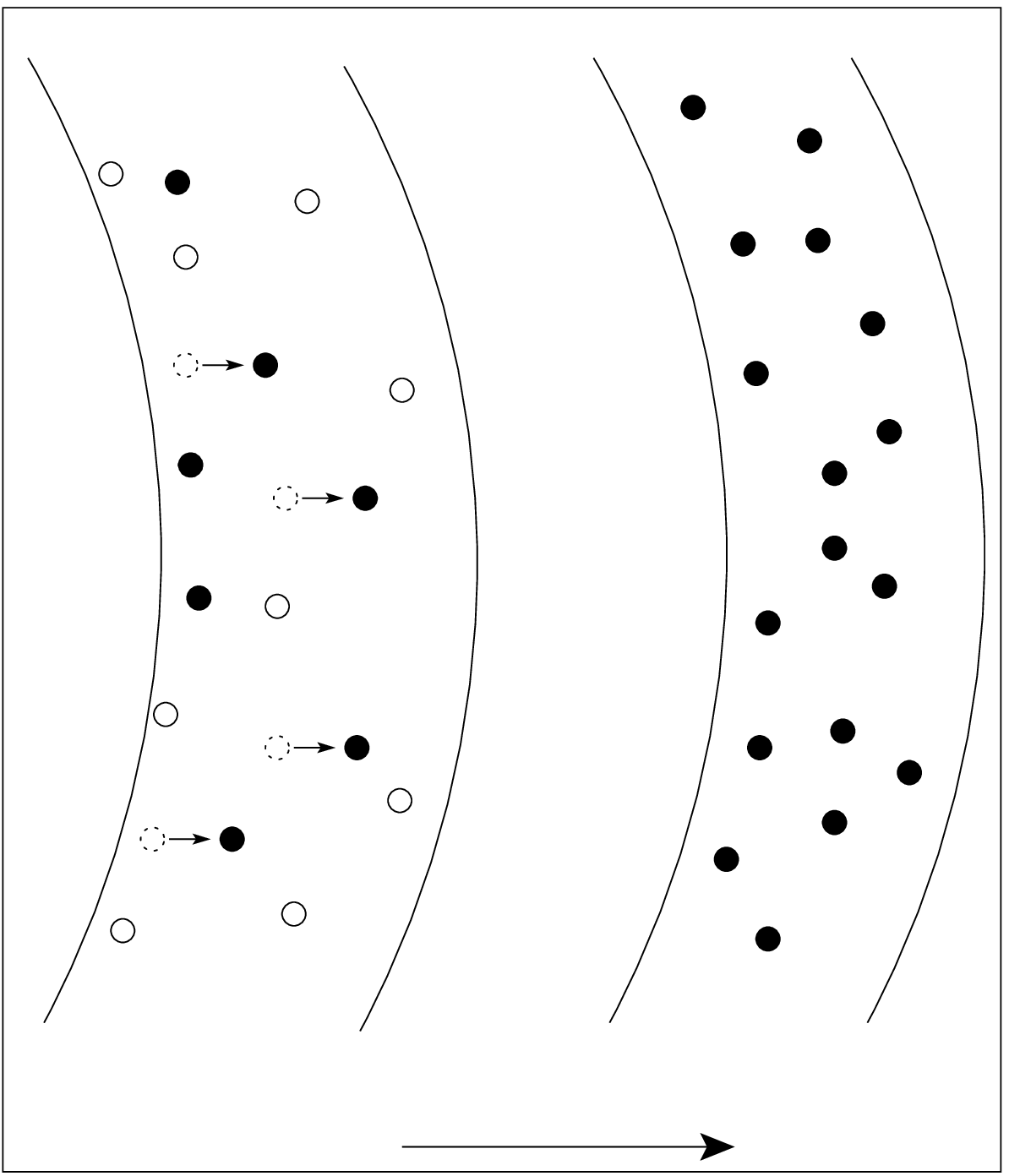}
\includegraphics[height=3in]{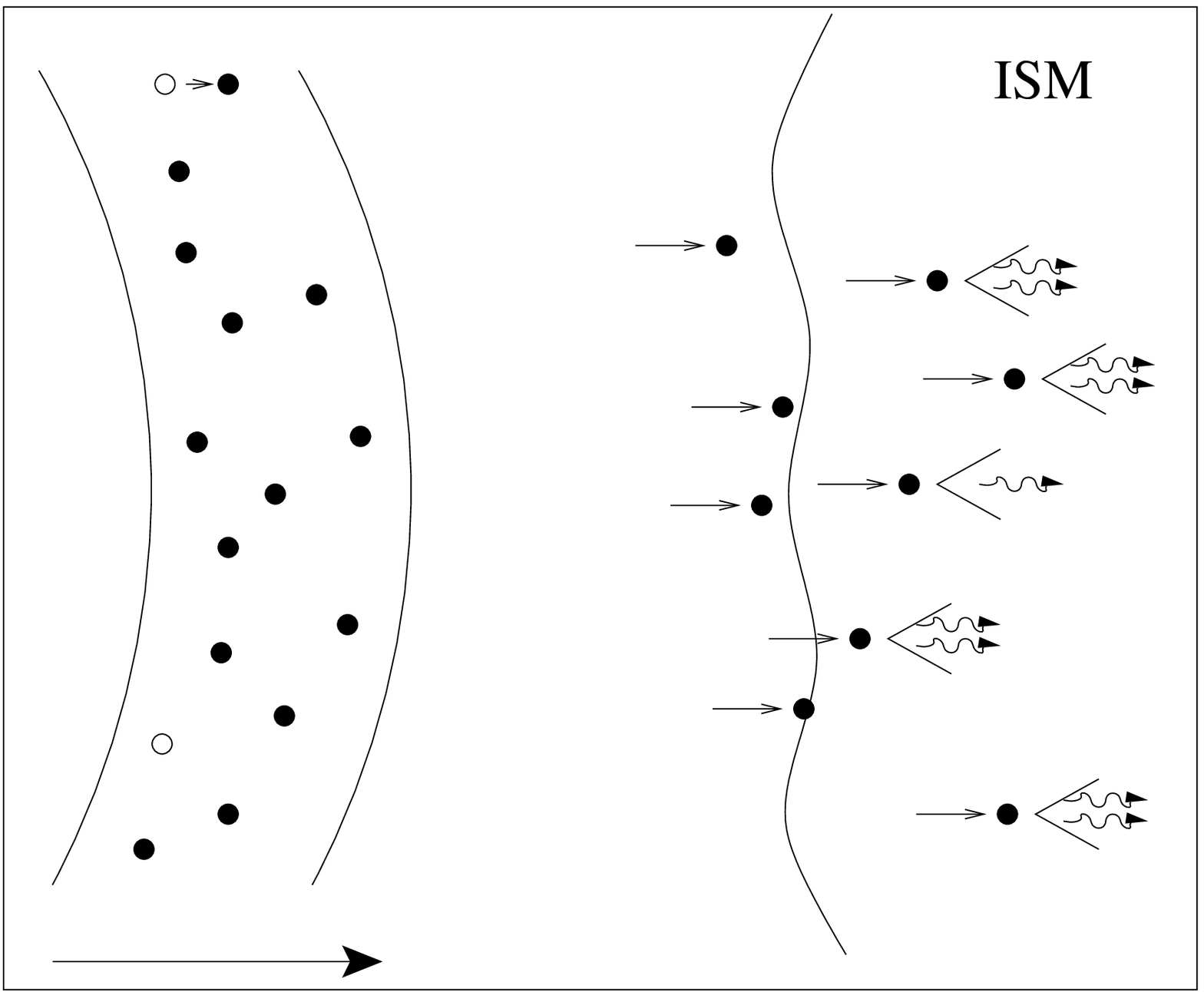}}
\caption{Illustration of the emission of a fireball by a compact
central engine (upper left), neutron decoupling (upper right), neutron 
decay (lower left), and the shocking of the outer shell with the 
ISM and 
imminent shocking of the slower neutron decayed shell
with the decelerating outer shell (lower right). The dark circles 
represent protons, the light circles neutrons, and the squiggly lines
represent the radition driving the acceleration 
of the fireball. \label{fig1}}
\end{figure}

\begin{figure}
\plotone{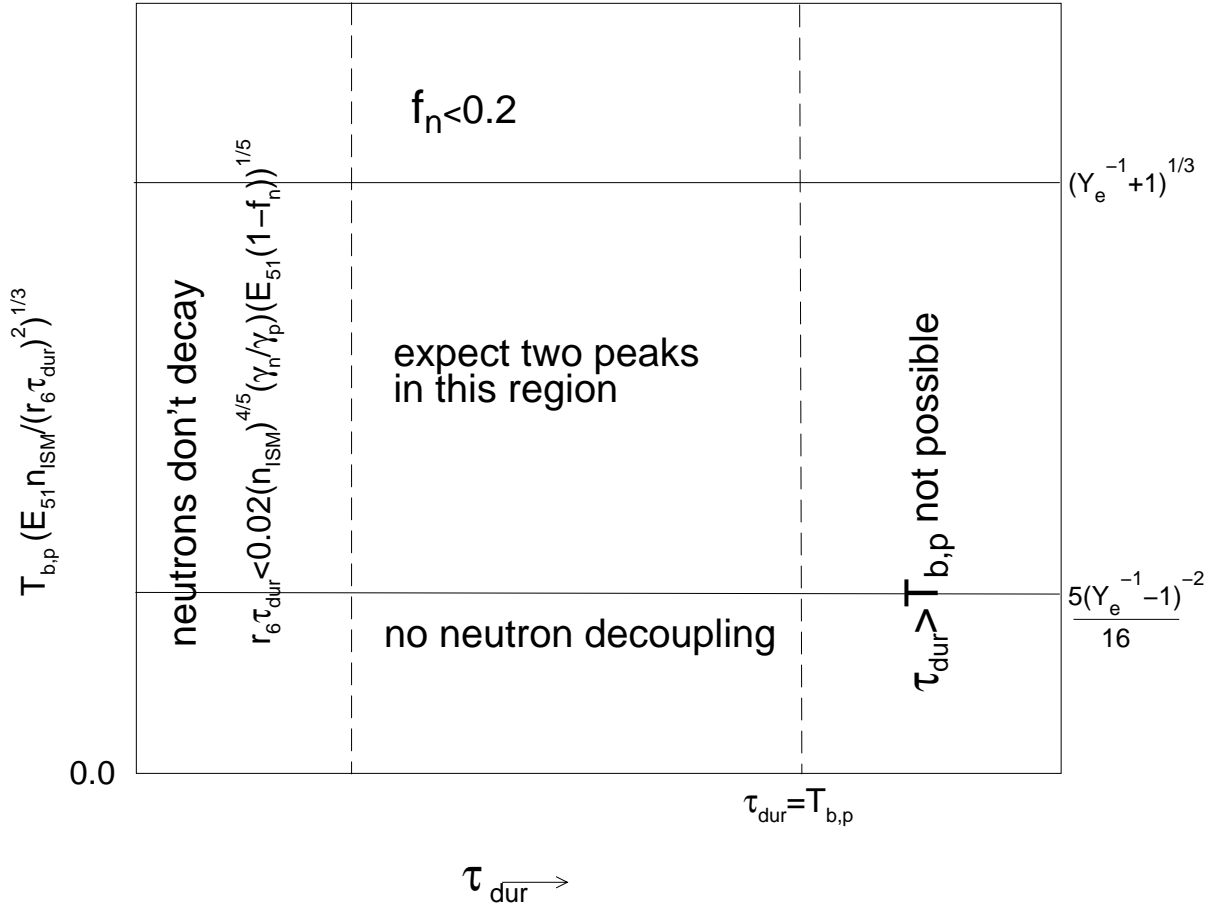}
\caption{Illustration of the region of the 
parameter space for which an observable second peak arising from the 
shocking of a decayed neutron shell with a decelerating outer shell 
is or is not expected.
The center box corresponds to the region 
for which one
expects the presence of a second peak when $\xi>1$ initially and
driven to unity through spreading. In the region to the left of the
central box the neutrons do not decay before colliding with the 
outer shell. Observable emission may result in this case but this possibility
has not been studied in detail.
The region above the central box corresponds to too 
little energy in the neutron component (neutron decoupling is too 
pronounced), while the region below the central box corresponds to 
no neutron decoupling. The region to the right of the central box
is not allowed because the observed burst duration is always at least
$\tau_{\rm dur}$. \label{fig2}}
\end{figure}

\begin{figure}
\plotone{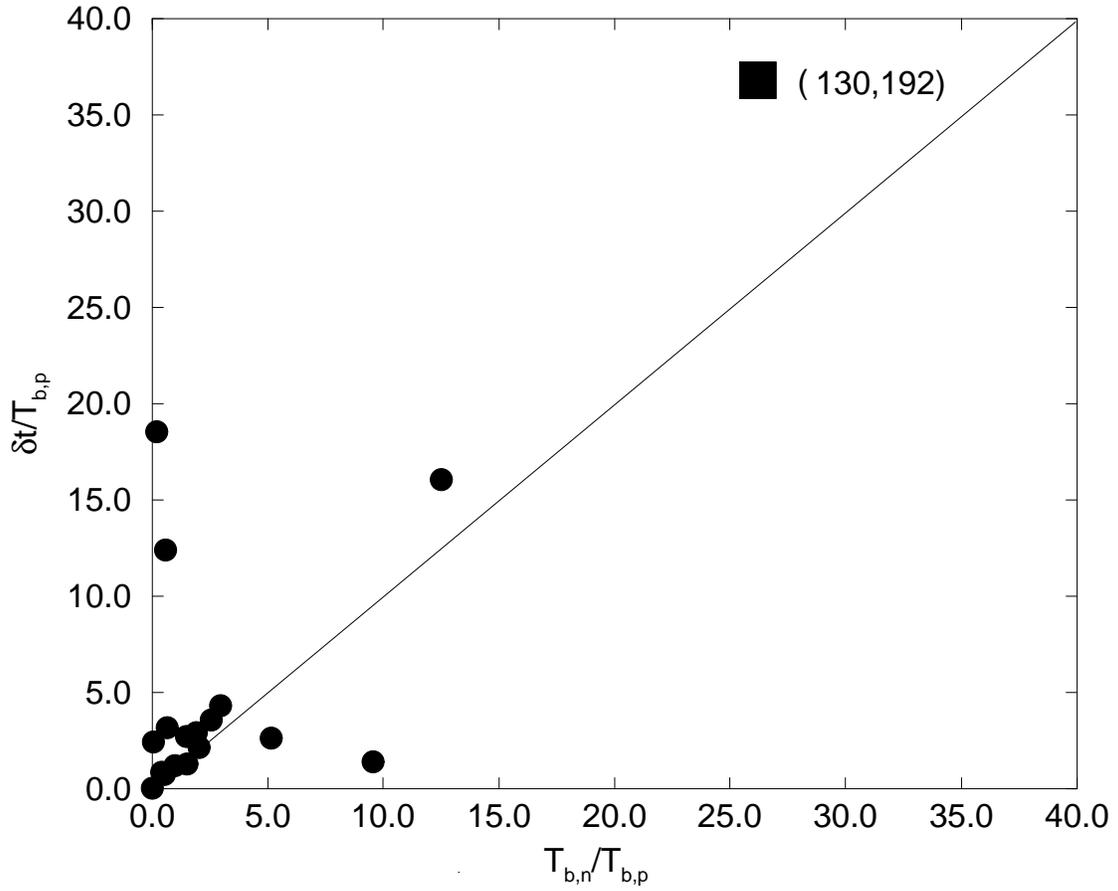}
\caption{Plot of $\delta t/T_{b,p}$ versus $T_{b,n}/T_{b,p}$ for the
bursts fit by Lee to have only two peaks in energy channel 2.  The box
in the upper right hand corner denotes a burst with $(\delta
t/T_{b,p},T_{b,n}/T_{b,p})=(130,192)$. Bursts falling near and to the
right of the line are consistent with the neutron decoupling picture.
\label{fig3}}
\end{figure}

\end{document}